\documentclass[conference]{IEEEtran}
\IEEEoverridecommandlockouts
\usepackage{cite}
\usepackage{amsmath,amssymb,amsfonts}
\usepackage{bbm}
\usepackage{algorithmic}
\usepackage{graphicx}
\usepackage{textcomp}
\usepackage{xcolor}
\usepackage{dsfont}
\usepackage{physics}
\usepackage{comment}
\usepackage{hyperref}
\usepackage[acronym]{glossaries}
\def\BibTeX{{\rm B\kern-.05em{\sc i\kern-.025em b}\kern-.08em
    T\kern-.1667em\lower.7ex\hbox{E}\kern-.125emX}}

\counterwithin{equation}{section}
\newcounter{thm}
\newtheorem{theorem}[thm]{Theorem}

\newtheorem{definition}[thm]{Definition}

\newtheorem{lemma}[thm]{Lemma}

\newacronym{bpsk}{BPSK}{Binary Phase Shift Keying}
\newacronym{POVM}{POVM}{Positive Operator Valued Measurement}

\begin{document}

\title{Private Communication over a Bosonic Compound Channel\\

\thanks{This work was financed by the DFG via grant NO 1129/2-1 and by the BMBF via grants 16KISQ039 and 16KISQ077. The authors acknowledge the financial support by the Federal Ministry of Education and Research of Germany in the programme of “Souverän. Digital. Vernetzt.”. Joint project 6G-life, project identification number: 16KISK002. M.R. acknowledges support from the project PNRR Id MSCA\_0000011-SQUID - CUP F83C22002390007 (Young Researchers).}
}

\author{\IEEEauthorblockN{Florian Seitz, Matteo Rosati, \'Angeles Vázquez-Castro, Janis Nötzel}
\textit{Technical University of Munich, Munich, Germany, \{flo.seitz,janis.noetzel\}@tum.de}\\
\textit{DICITA, Università Roma Tre, Via Vito Volterra 62, I-00146 Rome, Italy, matteo.rosati@uniroma3.it}\\
\textit{Autonomous University of Barcelona, Institut of Space Studies of Catalonia, Barcelona, Spain,}\\
\textit{angeles.vazquez@uab.cat}\\
}

\maketitle

\begin{abstract}
It is a common belief that quantum key distribution systems are the one and only information-theoretically secure physical layer security protocol that enables secure data transmission without a need for the legitimate parties to have any channel knowledge. It is also known that this high security profile results in severe rate restrictions for the parties utilizing the quantum key distribution systems. This observation raises the question of whether quantifying the level of ignorance of the legitimate parties with regard to the channel parameters may enable us to navigate the large gray zone between insecure but highly performant systems on the one side and perfectly secure but highly non-performant systems on the other side. Indeed, by proving a capacity formula for the bosonic compound wiretap channel using the binary phase shift keying alphabet, we are able to quantify in this work exactly how channel uncertainty penalizes data transmission rates.
\end{abstract}

\section{Introduction}
A compound channel models data transmission over a noisy channel when the channel parameters, such as loss or the variance of additive noise, are not completely known to the sending and receiving party. Rather, what is given to them are bounds on said parameters and a guarantee that the true parameter exists and is within those bounds.\\
Data transmission over compound channels has been studied as early as 1959 by Blackwell, Breiman, and Thomasian \cite{bbt59}. For data transmission over quantum channels, first coding theorems for finite-dimensional classical-quantum compound channels have been proven independently in \cite{bb2009,hayashi2009}. A highly efficient approach to proving compound capacity formulae for quantum channels has been demonstrated in \cite{bbjn}. In \cite{cnr}, a first step was taken towards understanding some practically relevant bosonic compound channels.\\
It may be argued that the unknown channel parameters in the bosonic compound model, such as transmissivity or noise variance, could be measured by the legal parties. This reasoning does however not apply to the parameters describing information leakage to an eavesdropper. For the link between the legal parties, the feedback channel will impact transmission latency. An initial discussion of the interplay between communication latency and channel uncertainty can be found in \cite{seitzNoetzel}.
\begin{figure}
    \centering
    \includegraphics[width=.4\textwidth]{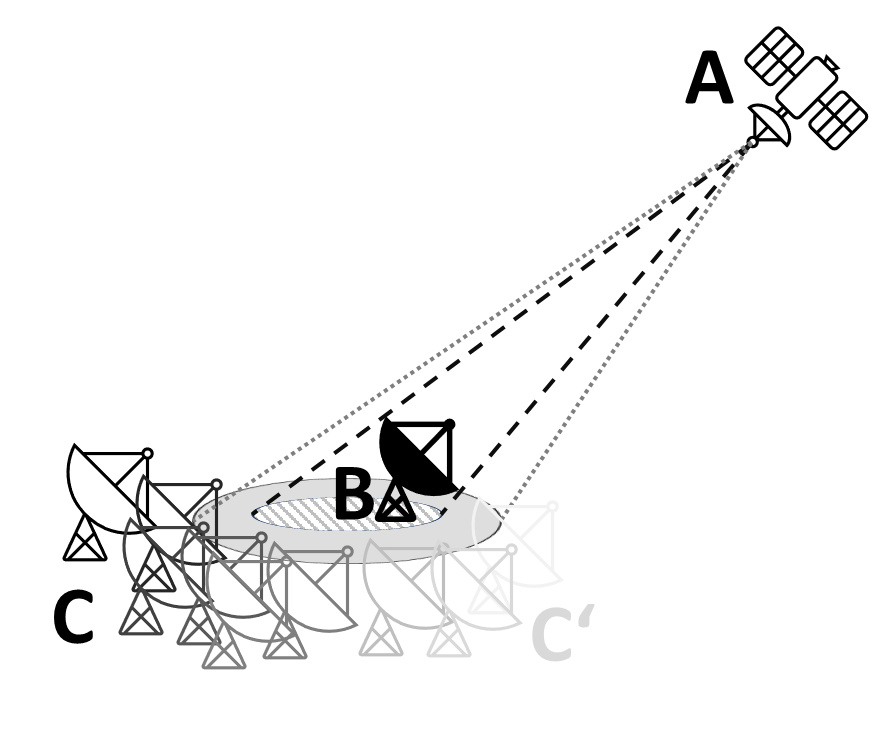}
    \caption{Satellite downlink with legal transmitter A and legal receiver B which do not know the exact location of an eavesdropper (reflected by two different possible positions C and C'), but can ensure it is not within a certain region around the legal receiver (striped region around B). Due to pointing errors (solid gray regions), variations in the channel transmissivity occur which result in a large region of possible parameters for the transmission lines A$\to$C or A$\to$C' or any other possible connection from A to a different wiretapper C''. Thus the actual exclusion zone is strictly larger than the striped region.}
    \label{fig:setup}
\end{figure}
The focus of this contribution is instead on secure communication under channel uncertainty. Focusing on the security aspect as the core problem, we assume the transmitting party A and to have perfect channel state information about the channel A$\to$B connecting it to the legal receiver B. On the contrary, since an eavesdropper cannot be expected to collaborate with the legal parties to help them gain channel information, it must be assumed that there is some degree of uncertainty that the legal communicating parties have with regard to the channel connecting the wiretapper to the legal transmitter. They must therefore prepare for a large class of possible channel parameters to arrive at a security guarantee. To avoid making assumptions on the hardware used by the wiretapper, they have to even assume perfect channel state information of the link A$\to$C at the wiretapper.

With this contribution, we first provide an analytical formula for the compound quantum channel capacity in the described setting under the assumption of \gls{bpsk} as the modulation format. We then describe a satellite downlink scenario following the derivation of \cite{angeles}, which motivates specific parameter choices and shows how the corresponding data rates relate to comparable systems.

In order to guarantee a positive data transmission rate in the system under study, the propagation characteristics of the laser beam used for transmission of the data from A to B were investigated in \cite{angeles}, and a geometric criterion was derived according to which secure communication is possible as long as it can be guaranteed that no wiretapper enters a specific exclusion zone (striped region around B in Figure \ref{fig:setup}). While the previous work \cite{angeles} provided security arguments based on entropic consideration, our contribution incorporates these already at the level of operational definition, allowing us to quantify security guarantees similar to the one of \cite{angeles} in generic situations of incomplete information.

For completeness, we point out that the concept of Wyner's wiretap channel \cite{wyner} has been investigated from a classical data transmission perspective over several decades, with recent efforts even leading to experimental demonstrations \cite{securePhyUplink}. 

\section{Notation and System Model}
The Fock space is denoted $\mathcal H$, the set of all density operators on it is $\mathcal{P}(\mathcal{H})$. The von-Neumann Entropy is written as $S$, the binary entropy is $h:[0,1] \to [0,1]; x \mapsto -x\log x - (1-x) \log(1-x)$ and $\overline{H}(x^n) = -\frac{1}{n} \log p_{X^n}(x^n)$ is the sample entropy of a sequence $x^n$. The uniform distribution on a set $\mathcal U$ is $\pi_\mathcal U$.
The entropy of the \gls{bpsk} ensemble is $h_{\scriptsize\scriptscriptstyle{BPSK}}(x):=h(\cosh(x)e^{-x})$~\cite{Guha2011,Rosati2023}.
The Holevo quantity of an ensemble $\{p_i,\rho_i\}_i$ is $\chi(\{p_i,\rho_i\}_i)$.
In this work, we focus on noiseless bosonic classical-quantum channels which are defined by one parameter $\tau\in[0,1]$, where $\tau^2$ is called the transmissivity and $1-\tau^2$ is the loss. Upon receiving the input $\alpha\in\mathbb C$, the channel output is given by $\mathcal N_\tau(\alpha)=|\tau\alpha\rangle\langle\tau\alpha|$. A noiseless bosonic compound wiretap channel is then given by a set $\mathcal N=\{\mathcal N_{\tau},\mathcal 
 N_{\eta}\}_{(\tau,\eta)\in\mathcal S}$ of channels where $\mathcal S$ is a subset of $[0,1]\times[0,1]$, $\tau$ quantifies the transmissivity of the transmission link between the legal receivers and $\eta$ that of the link from the legal sender to the wiretapper. The set $\mathcal S_B:=\{\tau:(\tau,\eta)\in\mathcal S\ \mathrm{for\ some\ }\eta\}$ and the set $\mathcal S_E:=\{\eta:(\tau,\eta)\in\mathcal S\ \mathrm{for\ some\ }\tau\}$ specify the individual links.
\begin{definition}[$(n,M_n,\lambda,\mu)$ Code] 
    A finite collection $\{x_{\tau,m}\}_{m\in[M_n],\tau\in\mathcal S_B}\subset\mathcal X^n$ of signals and a \gls{POVM} $\{D_m\}_{m=1}^{M_n}$ are called an $(n,M_n,\lambda,\mu)$ code $\mathcal C$ with state information at the transmitter if the success probability 
    \begin{align}\label{eq:p_suc_def}
        p_{\mathrm{suc}}(\mathcal C):=\min_{\tau\in\mathcal S_B}\frac{1}{M_n}\sum_{m=1}^{M_n}\Tr[D_m \mathcal N_\tau^{\otimes n}(x_{\tau,m})]
    \end{align}
    satisfies $p_\mathrm{suc}(\mathcal C)\geq1-\lambda$ and the information leakage to the eavesdropper satisfies
    \begin{align} \label{eq:privacy}
        \sup_{(\tau,\eta)\in\mathcal S}\chi(\{M_n^{-1},N_\eta^{\otimes n}(|x_{\tau,m}\rangle\langle x_{\tau,m}|)\}_{x_m=1}^{M_n}) < \mu. 
    \end{align}
    If each symbol $x_{m,,\tau,i}$ respects the constraint $x_{m,\tau,i}\in\mathcal R$ for some finite set $\mathcal R\subset\mathbb C$, then the code is said to use a modulation format. In particular, if $\mathcal R=\{-\sqrt{E},\sqrt{E}\}$ then the code is said to use \gls{bpsk} with energy $E$.
\end{definition}

\begin{definition}[Achievable Rates, Capacity]
    A rate $R\geq0$ is called achievable for the classical-quantum compound wiretap channel $\mathcal N$ under state constraint $\mathcal R$ if there exists a sequence $(\mathcal C_n)_{n\in\mathbb N}$ of $(n,M_n,\lambda_n,\mu_n)$ codes with state information at the transmitter, obeying the state constraint $\mathcal R$, such that both $\lambda_n\to0$ and $\mu_n\to0$ and $\limsup_{n\to\infty}\frac{1}{n}\log M_n\geq R$. 
    
    The secret message transmission capacity of $\mathcal N$ with state information at the transmitter is then defined as the supremum over all rates that are achievable for $\mathcal N$ and is denoted as $C_{ST}(\mathcal N)$.
\end{definition}
\section{Results}
    \subsection{Capacity Formula}
        Our results concern the specific compound channel arising from the restriction of \gls{bpsk} modulation with constraint $\mathcal R=\{-\sqrt{E},\sqrt{E}\}$ imposed on the signal states.
            \begin{theorem}
                For the compound channel defined by the state set $\{(\tau,\eta)\}_{\eta\in\mathcal T}$ where $\mathcal T\subset[0,1]$ and the state set $\mathcal R=\{-\sqrt{E},\sqrt{E}\}$ it holds hat
                \begin{align}
                    C = \min_{\tau\in\mathcal S_B}h_{\scriptscriptstyle{BPSK}}(\tau E) - \max_{\eta\in\mathcal S_E}h_{\scriptscriptstyle{BPSK}}(\eta E).\label{eqn:capacity-formula}
                \end{align} 
            \end{theorem}
        We note that, despite the simplifying assumption of a known fixed value of $\tau$, our results are nontrivial in the sense that they guarantee the existence of near-optimal codes which are secure for all possible values $\eta\in\mathcal T$.
    \subsection{Satellite Channel Model}
\begin{figure}[ht]
            \centering
            \includegraphics[width=0.45\textwidth]{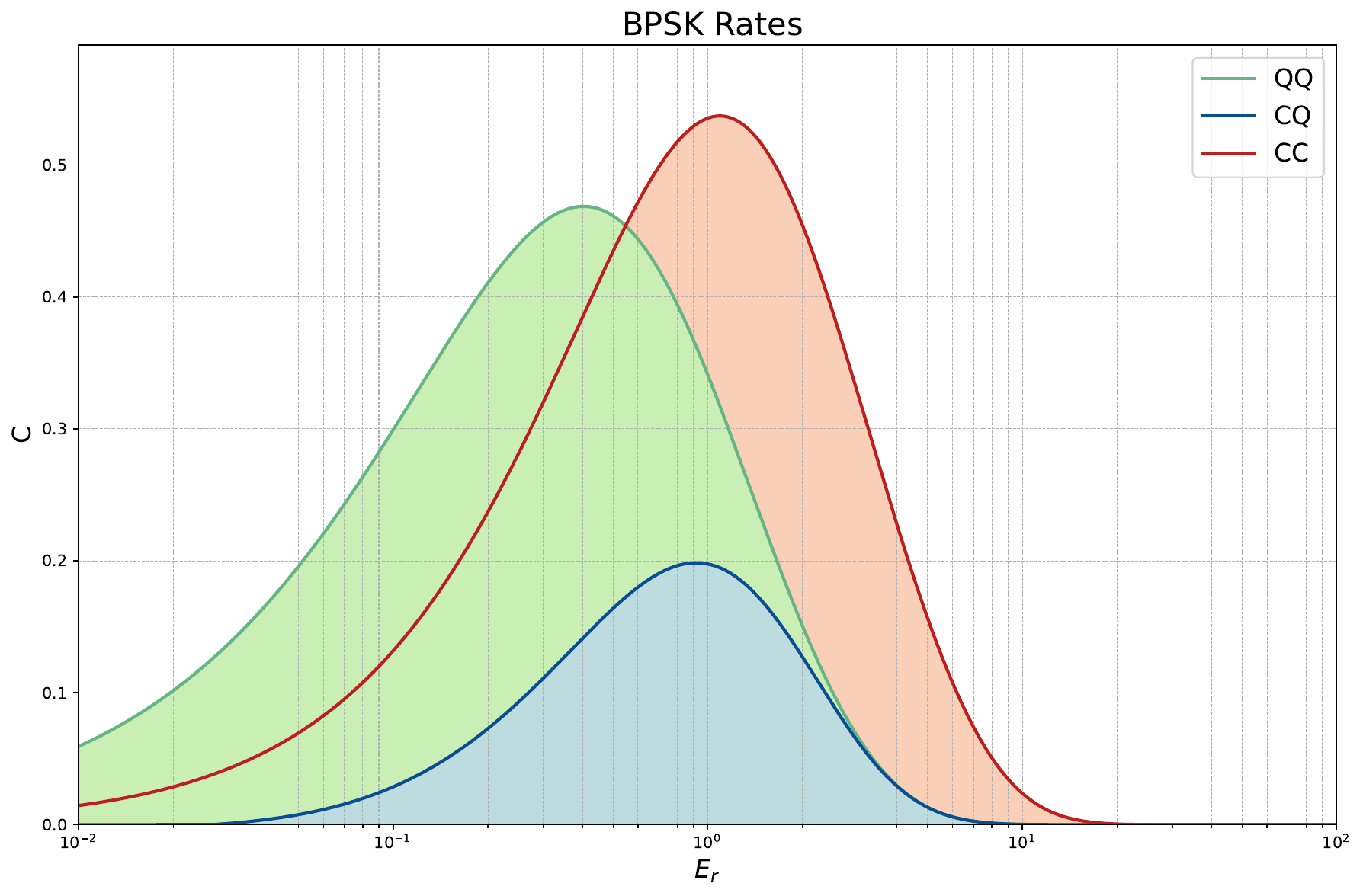}
            \caption{Comparison of private capacities attainable on the compound channel described in the text; when \eqref{eqn:capacity-formula} or \eqref{eqn:homodyne_private_capacity} become negative, the capacity value is taken to be zero. CQ refers to a strategy with coherent homodyne detection at the legitimate receiver, while QQ corresponds to a strategy with quantum-optimal multi-symbol detection, and CC refers to a situation where both the legitimate receiver and the wiretapper are limited to homodyne detection. The corresponding capacity, shown in red, falsely predicts secure communication rates, assuming that all parties are limited to homodyne detection. The blue region corresponds to a worst case scenario where only the wiretapper has access to a quantum receiver. Conversely, the green region shows the security gain achieved when the legitimate receiver also uses quantum technology, underscoring the importance of a quantum receiver when a potential adversary may have quantum capabilities.}
            \label{fig:plot}
        \end{figure}

        We utilize the satellite downlink analysis provided in \cite{satelliteToGround} to derive explicit values for the variability of $\eta$, arriving at a fluctuation of one order of magnitude over a duration of $12\mathrm{s}$. These values are taken from an actual satellite downlink experiment with a sampling rate of $20\mathrm{khz}$, where the distance between satellite and receiver was in the order of $750\mathrm{km}$. Assuming the transmissivity is constant during
        periods of $10^{-2}$ seconds, the receiver telescope will have enough time to transmit channel state information to the satellite. This leaves a time window of $10^{-2}/2$ seconds for data transmission, which a standard communication channel operating at $10\mathrm{GBd}$ could utilize to transmit $10^{8}/2$ symbols.
        Therefore, we assume our asymptotic formula \eqref{eqn:capacity-formula} to capture the central aspects of the satellite downlink scenario in this particular situation.
        To demonstrate the importance of formula \eqref{eqn:capacity-formula} we contrast the values predicted by it with formulas which we \emph{expect} to hold when the receivers have to obey certain restrictions in their decoding process:

        We compare the \gls{bpsk} private capacity attainable via general quantum coding and decoding operations \eqref{eqn:capacity-formula} (dubbed QQ), with that attainable when the legitimate receiver employs a classical coherent homodyne detector to decode the phase of the received state. For the sake of comparison we assume, analogously to \eqref{eqn:capacity-formula}, that that the compound secrecy capacity is given by the smallest secrecy capacity of the individual channels. This can be obtained by computing the Shannon capacity of a binary-symmetric channel with error probability for states of mean photon number $E$ given by
        \begin{equation}
            P_E = \frac12\left(1 - {\rm Erf}\left(\sqrt{2 E}\right)\right),
        \end{equation}
        resulting in a private capacity (dubbed CQ)
        \begin{equation}\label{eqn:homodyne_private_capacity}
            C_{\scriptscriptstyle{BPSK-CQ}} = 1-\max_{\tau\in\mathcal S_B}h\left(P_{\tau E}\right) - \max_{\eta\in\mathcal S_E}h_{\scriptscriptstyle{BPSK}}(\eta E).
        \end{equation}
        For comparison, a private capacity where both the legitimate receiver and the wiretapper use homodyne detection is given as 
        \begin{equation}\label{eqn:hh_private_capacity}
            C_{\scriptscriptstyle{BPSK-CC}} = 1-\max_{\tau\in\mathcal S_B}h\left(P_{\tau E}\right) - \max_{\eta\in\mathcal T}\left(P_{\eta E})\right).
        \end{equation}
        In order to outline the relevance of the analysis we compare in Fig.~\ref{fig:plot} the private capacities as a function of photon-number at the receiver $E_r = \tau^2 E$, for fixed $E=10^6$ modeling a $0.1\mathrm{W}$ transmitter and $\eta_{\max}^2\in[0.02\cdot\tau^2,\tau^2\cdot 0.2]$, where in formulas \eqref{eqn:homodyne_private_capacity} and \eqref{eqn:hh_private_capacity} we pick the worst-case value $\eta^2=\tau^2\cdot 0.2]$ for the wiretapper and we consider a range $\tau \in [10^{-2},10^{-4}]$.

\section{Preliminaries}
In this part, we introduce concepts from the literature that we are going to use to derive our main result. The packing and covering lemmas can be found in \cite{wilde2017} (chapters 16 and 17).

\begin{lemma}[Covering Lemma]
Let $\mathcal X$ be a set and $p_{X}(x)$ a probability distribution over $\mathcal X$. Then $\left\{ p_{X}(x), \rho_{x} \right\}_{x \in \mathcal X}$, where $\left\{\rho_{x}\right\}_{x \in \mathcal X} \subset \mathcal P(\mathcal H)$ for some $\mathcal{H}$, is called the true ensemble. Let $\mathcal L$ be a set and select a random code word $X_l$ according to $p_{X}(x)$ for each $l \in \mathcal L$, then $\left\{ \frac{1}{\left| \mathcal L \right|},\rho_{X_l} \right\}_{l \in \mathcal L}$ is called the fake ensemble. Let there be a code space projector $\Pi$, that is an orthogonal projector with the properties
\begin{align}
\mathrm{Tr}\left[\rho_{x} \Pi \right] &\geq 1-\varepsilon, \\
\mathrm{Tr}\left[\Pi\right] &\leq D,
\end{align}
and a set of codeword projectors $\left\{ \Pi_{x} \right\}_{x \in \mathcal X}$ with the properties
\begin{align}
\mathrm{Tr}\left[\rho_{x}\Pi_{x}\right] &\geq 1-\varepsilon, \\
\Pi_{x} \rho_{x}\Pi_{x} &\leq \frac{1}{d}\Pi_{x},
\end{align}where $\varepsilon > 0$ and $0<d<D$. Let $\bar{\rho} = \sum_{x \in \mathcal X}p_{X}(x) \rho_{x}$ be the average state of the true ensemble and let $\bar{\rho}_{\mathcal L} = \frac{1}{|\mathcal L|}\sum_{l \in \mathcal L} \rho_{s}$ be the average state of the fake ensemble. Then
\begin{align}
\mathrm{Pr}\left\{ \|\bar{\rho}-\bar{\rho}_{\mathcal L}||_{1} \leq 30\varepsilon^{\frac{1}{4}} \right\} &\geq 1-2D \exp\left( -\frac{\varepsilon^{3}|\mathcal L|d}{4 D} \right),
\end{align}
given $\varepsilon$ is small and $|\mathcal L| \gg \frac{\varepsilon^{3}d}{D}$. The probability is with respect to the random code book selection. Note that the original bound is a bit tighter.
\end{lemma}
\begin{lemma}[Packing Lemma]
Let $\mathcal X$ be a set and $p_{X}(x)$ a probability distribution over $\mathcal X$. Then $\left\{ p_{X}(x), \rho_{x} \right\}_{x \in \mathcal X}$, where $\left\{\rho_{x}\right\}_{x \in \mathcal X} \subset \mathcal P(\mathcal H)$ for some $\mathcal{H}$, is an ensemble with the average state $\bar{\rho} = \sum_{x \in \mathcal X}p_{X}(x) \rho_{x}$. Suppose there exists a code space projector $\Pi$ with the properties
\begin{align}
    \Tr[\Pi \rho_x] &\geq 1-\varepsilon,\\
    \Pi \bar{\rho} \Pi &\leq \frac{1}{D} \Pi,
\end{align}
and a set of code word projectors $\left\{ \Pi_x \right\}_{x \in \mathcal X}$ with the properties
\begin{align}
    \Tr[\Pi_x \rho_x] &\geq 1-\varepsilon,\\
    \Tr[\Pi_x] &\leq d,
\end{align}
for some $\varepsilon \in (0,1)$ and $0<d<D$. Let $\mathcal M$ be a set of messages and let $\mathcal C = \left\{C_m\right\}_{m \in \mathcal M}$ be a random code, where every element $C_m$ is selected randomly and independently from $\mathcal X$ according to $p_X(x)$. Then there exists a \gls{POVM} $\left\{\Lambda_m\right\}_{m \in \mathcal M}$ for which the expected probability of distinguishing the states $\left\{ \rho_{C_m} \right\}_{m \in \mathcal M}$ is high:
\begin{align}
    \mathbb{E}_{\mathcal{C}}  \frac{1}{|\mathcal{M}|} \sum_{m \in \mathcal{M}} \Tr[ \Lambda_m \rho_{C_m}  ] \geq 1 - 6 \sqrt{\varepsilon} - 4 |\mathcal{M}| \frac{d}{D},
\end{align}
where $\mathds E_{\mathcal C}$ denotes the expectation value with respect to the random variable $\mathcal C$, $\frac{D}{d}$ is large, $\abs{\mathcal M} \ll \frac{D}{d}$ and $\varepsilon \ll 1$. The original bound is again a bit tighter.
\end{lemma}

\begin{lemma}[Finite Support Approximation] \label{thm:fin_sup}\\
    Let $0 \leq \rho, \sigma, \Lambda \leq \mathds{1}$. Then
    \begin{align}
    \Tr[\Lambda \rho] \leq \Tr[\Lambda \sigma] + \| \rho - \sigma \|_1.
    \end{align}
A proof is given in \cite{wilde2012}.
\end{lemma}
\begin{lemma}[Von-Neumann Entropy Continuity Bound]\label{thm:entropy_continuity}
Let $\hat{N}$ denote the photon number operator, and consider two quantum states $\rho$ and $\sigma$ on an infinite-dimensional separable Hilbert space $\mathcal{H}$. Suppose these states satisfy the energy constraints
\begin{align}
\mathrm{Tr}\left[\hat{N} \rho\right], \mathrm{Tr}\left[\hat{N} \sigma\right] \leq E \leq \infty,
\end{align}
and are close in trace distance:
\begin{align}
\frac{1}{2} \left\| \rho - \sigma \right\|_1 \leq \varepsilon,
\end{align}
where $0 \leq \varepsilon \leq \frac{E}{1 + E}$. Under these conditions, the von Neumann entropies of $\rho$ and $\sigma$ satisfy the continuity bound
\begin{align}
\left| S(\rho) - S(\sigma) \right| \leq h(\varepsilon) + E \cdot h\left( \frac{\varepsilon}{E} \right).
\end{align}
The version is from \cite{Becker2021}, based on \cite{Winter2015}.
\end{lemma}
We also make extensive use of the concepts of quantum typicality and strong classical typicality. Here, we only introduce some key properties; for a full discussion, refer to \cite{wilde2017}.
\begin{definition}[Strong Typicality]
Let $\mathcal X$ be a finite set, $x^n = (x_1, \dots, x_n)$ a sequence where each $x_i$ is an element of $\mathcal X$ and $N(x|x^n)$ the number of times the symbol $x$ occurs in $x^n$. Let $p_X(x)$ be a probability distribution for a random variable $X$. Then for $\delta > 0$, the $\delta$-strongly typical set $T_\delta^{X^n}$ consists of all sequences $x^n$ where
\begin{align}
    \forall x \in \mathcal{X}, 
\begin{cases}
\left| \frac{1}{n} N(x|x^n) - p_X(x) \right| \leq \delta & \text{if } p_X(x) > 0, \\
\frac{1}{n} N(x|x^n) = 0 & \text{else}
\end{cases}.
\end{align}
For $\varepsilon, \delta > 0$, sufficiently large $n$ and a positive constant $c$, $T_{\delta}^{X^n}$ has the properties
\begin{align}
    \text{Pr}\left\{X^n \in T_\delta^{X^n}\right\} &\geq 1-\varepsilon\\
    (1-\varepsilon) 2^{n (H(X) + c \delta)} &\leq \abs{T_\delta^{X^n}} \leq 2^{n (H(X) + c \delta)}
\end{align}
\end{definition}
\begin{definition}[Typical Subspace]
Let $\rho \in \mathcal{P}(\mathcal{H})$ be a density operator with the spectral decomposition $\rho = \sum_{x=0}^\infty p_X(x) \ket{x}\bra{x}$. For every $n \in \mathbb{N}$, the projector onto the typical subspace of $\rho^{\otimes n}$ is defined as
\begin{align}
\Pi_{\rho, \delta}^n = \sum_{x^n \in T_\delta^{X^n}} \ket{x^n}\bra{x^n},
\end{align}
where $T_\delta^{X^n}$ is the set of typical sequences of length $n$ for $p_{X^n}(x^n) = p_X(x_1) \dots p_X(x_n)$.
It has the following properties for sufficiently large $n$ and any $\epsilon, \delta > 0$:
\begin{align}
\mathrm{Tr}\left[\Pi_{\rho, \delta}^n\right] &\leq 2^{n \left( S(\rho) + \delta \right)}, \\
\mathrm{Tr}\left[\Pi_{\rho, \delta}^n \rho^{\otimes n}\right] &\geq 1 - \epsilon, \\
2^{-n\left( S(\rho) + \delta \right)} \Pi_{\rho, \delta}^n &\leq \Pi_{\rho, \delta}^n \rho^{\otimes n} \Pi_{\rho, \delta}^n \leq 2^{-n\left( S(\rho) - \delta \right)} \Pi_{\rho, \delta}^n.
\end{align}
The subspace $\mathrm{span}\left\{ \ket{x^n} : x^n \in V_\delta^n \right\}$ is referred to as the weakly Typical Subspace. This subspace captures the majority of the probability mass of the state $\rho^{\otimes n}$.
\end{definition}
\begin{lemma}[Coherent State Approximation]\label{thm:coher_approx}
Let $\ket{\alpha}$ be a coherent state and let $P_{N} = \sum_{n=0}^N \ket{n}\bra{n}$, where $\ket{n}$ are photon number states, be the projector onto a Fock space that is truncated at $N$. Then
\begin{align}
\mathrm{Tr}\left[P_{N} \ket{\alpha} \bra{\alpha} \right] \geq 1-\frac{1}{2} 2^{-N}
\end{align}
for all $N > 8e|\alpha|^2$. Clearly the same bound applies to a state $\rho=\sum_{\alpha \in A} p(\alpha) \ket{\alpha}\bra{\alpha}$ as long as $N > 8e|\alpha|^2 \; \forall \alpha \in A$.
\end{lemma}
Lemma \ref{thm:coher_approx} follows from Stirling's formula and the observation that for every $c>0$ we have $(c/N)^N<2^{-N}$ once $N$ is large enough.
\begin{lemma}[BPSK Ensemble Holevo Quantity] \label{thm:bpsk_chi}
Let $\rho_0:=\ket{\alpha}\bra{\alpha}$ and $\rho_1:=\ket{-\alpha}\bra{-\alpha}$ and without loss of generality $\alpha>0$. Then the Holevo quantity of the corresponding ensemble is~\cite{Guha2011,Rosati2023} 
\begin{align}
    \chi\left(\{\tfrac{1}{2},\rho_i\}_{i=0}^1\right) = h\left(\cosh(\alpha^2)e^{-\alpha^2}\right).
\end{align}
\end{lemma}
\section{Sketch of Proof}
In order to show the achievable secrecy capacity of the \gls{bpsk} ensemble for the noiseless bosonic channel, we begin by considering a general ensemble before narrowing our focus to this specific case. The idea is to exploit the fact that the wiretapper channel has a higher loss than the legitimate channel. As a consequence, the maximum number of distinguishable states that the legitimate receiver can reliably identify is greater than the corresponding number for the wiretapper. Therefore, a ``true'' ensemble can be partitioned into a set of ``fake'' ensembles, one for each message, in such a way that the average state of each fake ensemble looks the same to the wiretapper while the receiver is able to distinguish every state in the true ensemble, ensuring that the secrecy of the communication is preserved. In order to send a certain message, the sender selects one of the code words from the fake ensemble corresponding to this message and sends it over the channel. The packing lemma is used to show that the receiver can identify a certain number of states, while the covering lemma is used to show that the wiretapper is unable to distinguish more than a certain number of states. For the converse part, we note that the compound capacity is equal to the minimal capacity over the compound set.
\section{Proof}
Let $\tau, \eta \in [0,1]$ be the loss of the receiver and the wiretapper channel, respectively. The parameter $\tau$ is arbitrary but fixed and we therefore omit it in the notation. Consider sequences $x^n = x_1,\dots,x_n$, where each $x_i$ takes values from a discrete set $\mathcal X \subset \mathds C$ Then with $n$ uses of the channel the receiver gets the state $\rho_{x^n} = \rho_{x_{1}} \otimes \dots \otimes \rho_{x_{n}}$ and the wiretapper gets $\tilde{\rho}^\eta_{x_{1}} \otimes \dots \otimes \tilde{\rho}^\eta_{x_{n}}$, where $\rho_{x} = \ket{\tau x}\bra{\tau x}$ and $\tilde{\rho}^\eta_{x} = \ket{\eta x}\bra{\eta x}$. Kets and Bras denote coherent states.

We consider finite dimensional approximations of these states using lemmas \ref{thm:fin_sup} and \ref{thm:coher_approx}. For a state $\rho$ we set $\rho' = P_{N} \rho P_{N} + \mathrm{Tr}\left[(\mathds 1-P_{N})\rho\right]\ket{0}\bra{0}$, in order to keep the state normalized. Based on this definition and lemma \ref{thm:coher_approx} we get $\|\rho-\rho'\|_{1}\leq 2^{-N}$ for sufficiently large $N$. For $n$-mode states, we have to multiply this bound by $n$, which implies that to get a vanishing error for large $n$, we need $N \sim \log(n)$, meaning that our $n$ mode approximated state has $\sim \log(n)^n$ dimensions. If we set for example $N = 2 \log(n)$, the maximum approximation error of an $n$ mode state is upper bounded by $\frac{1}{n}$ for sufficiently large $n$.

Let $X^n$ be a random variable with probability distribution $p_{X^n}(x^n) = p_{X}(x_{1})  \dots  p_{X}(x_{n})$. based on this we define the ''pruned`` distribution $p'_{X'^n}$ by setting the probabilities of all sequences that are not contained in $T_{\delta}^{X^n}$, the set of strongly delta typical sequences of $p_{X^n}$, to zero and re-normalizing:
\begin{align}
    p'_{X^n}(x^n) = 
\begin{cases}
\frac{1}{\xi} p_{X^n}(x^n) & : x^n \in T_{\delta}^{X^n} \\
0 & : x^n \notin T_{\delta}^{X^n},
\end{cases}
\end{align}
where $\xi = \sum_{x^n \in T_{\delta}^{X^n}} p_{X^n}(x^n)$.
The distributions are approximately close, in the sense that $\lim_{ n \to \infty }S(p'_{X'^n})=S(p_{X^n})$.

The idea is to use the packing and covering lemmas to show that there exists a good code for the finitely approximated states $\rho'_{x^n}, \tilde{\rho}'^\eta_{x^n}$ and then argue that the same code and operators perform almost as well with the original states. Let $\rho'^n := \mathds E_{X'^n} \rho'_{x^n}$ be the average state at the receiver, where $\mathds E$ denotes the expectation value. Then we construct a set of projectors, namely, for the packing lemma, we need a code space projector $\Pi^n$ and a set of code word projectors $\Pi_{x^n}^n$ with the properties
\begin{align}
\mathrm{Tr}\left[\Pi^n \rho'_{x^n}\right] &\geq 1- \varepsilon,\label{eq:pack1}\\
\mathrm{Tr}\left[\Pi^n_{x^n}\rho'_{x^n}\right] &\geq 1-\varepsilon,\label{eq:pack2}\\
\mathrm{Tr}\left[\Pi_{x^n}^n\right]&\leq d,\label{eq:pack3}\\
\Pi^n \rho'^{n} \Pi^n &\leq  \frac{1}{D} \Pi^n.\label{eq:pack4}
\end{align}
For the covering lemma, we need similar projectors $\tilde{\Pi}^n$ and $\tilde{\Pi}^n_{x^n}$ with the properties
\begin{align}
\mathrm{Tr}\left[\tilde{\Pi}^n \tilde{\rho'}^\eta_{x^n}\right] &\geq 1-\tilde{\varepsilon},\label{eq:cover1}\\
\mathrm{Tr}\left[\tilde{\Pi}^n_{x^n} \tilde{\rho'}^\eta_{x^n}\right] &\geq 1-\tilde{\varepsilon},\label{eq:cover2}\\
\mathrm{Tr}\left[\tilde{\Pi}^n\right] &\leq \tilde{D},\label{eq:cover3}\\
\tilde{\Pi}^n_{x^n}\tilde{\rho'}^\eta_{x^n} \tilde{\Pi}^n_{x^n} &\leq \frac{1}{\tilde{d}} \tilde{\Pi}^n_{x^n}.\label{eq:cover4}
\end{align}
For the code word projectors we use the fact that our states are pure states, therefore they are themselves rank-one projectors and we can set $\Pi_{x^n} = \rho_{x^n}$, $\tilde{\Pi}^n_{x^n} = \tilde{\rho}^\eta_{x^n}$, $d = \tilde{d} = 1$, fulfilling the properties \eqref{eq:pack2}, \eqref{eq:pack3}, \eqref{eq:cover2} and \eqref{eq:cover4}. For the code space projector, we use the strong typical subspace projectors corresponding to the average states $\sigma = \sum_{x \in \mathcal X}p_X(x) \rho'_x$ and $\tilde{\sigma}^\eta = \sum_{x \in \mathcal X}p_X(x) \tilde{\rho}'^\eta_x$ and set $\Pi^n = \Pi^n_{\sigma, \delta}$ and $\tilde{\Pi}^n = \Pi^n_{\sigma^\eta,\delta}$. This definition enables us to directly apply property 15.2.7 from \cite{wilde2017}, ensuring that \eqref{eq:cover1} and \eqref{eq:pack1} are satisfied. To fulfill \eqref{eq:cover3} we set $\tilde{D} = 2^{n(S(\tilde{\sigma}^\eta)+\delta)}$, due to the properties of the typical subspace projector. For the last property, \eqref{eq:pack4}, the analysis in \cite{wilde2017}, chapter 20.3.1 shows that
\begin{align}
    \Pi^n \rho'^{n} \Pi^n &\leq \frac{1}{1-\varepsilon'}2^{-n(S(\sigma)-c' \delta)} \Pi^n,
\end{align}
where $c'$ is a positive constant and we used the property $\text{Pr}\left\{X^n \in T_\delta^{X^n}\right\} \geq 1-\varepsilon'$. This allows us to set $D = (1-\varepsilon') 2^{n(S(\sigma)-c' \delta)}$

We have seen earlier that for sufficiently large $n$ we have $\|\sigma - \Bar{\rho} \|_1 \leq \frac{1}{n}$ and $\|\tilde{\sigma}^\eta - \tilde{\rho}^\eta \|_1 \leq \frac{1}{n}$, where $\Bar{\rho} = \mathds E_X \rho_X$, $\tilde{\rho}^\eta = \mathds E_X \tilde{\rho}^\eta_X$ are the single-mode average state in infinite dimensions. With lemma \ref{thm:entropy_continuity} and the input energy constraint $E$ we see that the entropy of the approximated state approaches the entropy of the original states, as expected:
\begin{align}
    S(\sigma) &\geq S(\Bar{\rho}) - h\left(\frac{1}{n}\right) - E h\left(\frac{1}{En}\right)\\
    S(\tilde{\sigma}^\eta) &\geq S(\tilde{\rho}^\eta) - h\left(\frac{1}{n}\right) - E h\left(\frac{1}{En}\right)
\end{align}

We can now construct a code that fulfills the requirements for data transmission and secrecy by selecting $\abs{\mathcal M} \abs{\mathcal L}$ random variables $X_{m,l}^n$ according to $p'_{X'^n}(x^n)$. The corresponding states at the receiver are $\left\{\rho_{X_{m,l}^n}\right\}_{m \in \mathcal M, l \in \mathcal L}$. Then due to the packing lemma and Markov's inequality the probability that there exists a \gls{POVM} $\left\{\Lambda_{m,l}\right\}_{m \in \mathcal M, l \in \mathcal L}$ such that
\begin{align}
    &\frac{1}{\abs{\mathcal M} \abs{\mathcal L}} \sum_{m \in \mathcal M, l \in \mathcal L} \Tr[\Lambda_{m,l} \rho'_{X_{m,l}^n}]  \\\geq &1-6\sqrt{\varepsilon}- \frac{4 \abs{\mathcal M} \abs{\mathcal L}}{1-\varepsilon'}2^{-n(S(\sigma)-c' \delta)}, \nonumber
\end{align}
and therefore with Lemma \ref{thm:fin_sup}
\begin{align}
    &\frac{1}{\abs{\mathcal M} \abs{\mathcal L}} \sum_{m \in \mathcal M, l \in \mathcal L} \Tr[\Lambda_{m,l} \rho_{X_{m,l}^n}]  \\\geq &1-6\sqrt{\varepsilon}- \frac{4 \abs{\mathcal M} \abs{\mathcal L}}{1-\varepsilon'}2^{-n(S(\sigma)-c' \delta)} - \frac{1}{n}, \nonumber
\end{align}
is exponentially large. As a consequence, if $\abs{\mathcal M} \abs{\mathcal L} < 2^{nS(\Bar{\rho})}$, the success probability is arbitrarily close to one for sufficiently large $n$.

For the privacy part, consider the fake ensembles at the wiretapper, $\left\{ \frac{1}{\abs{\mathcal L}}, \tilde{\rho}^\eta_{X_{m,l}^n} \right\}_{l \in \mathcal L}$ for every $m \in \mathcal M$, with average states $\tilde{\rho}_{\eta,m}^n$. Then the probability with respect to the code construction that the average state of a fake ensemble is close in trace distance to the average state of the true ensemble is
\begin{align}
    \mathrm{Pr}\left\{ \|\tilde{\rho}_{\eta}^n-\tilde{\rho}_{\eta,m}^n||_{1} \leq 30\varepsilon^{\frac{1}{4}} + \frac{1}{n} \right\} &\geq 1-2\tilde{D} \exp\left( -\frac{\varepsilon^{3}|\mathcal L|}{4 \tilde{D}} \right).
\end{align}
Together with lemma \ref{thm:entropy_continuity}, this ensures that the information leakage to the eavesdropper \eqref{eq:privacy} becomes arbitrarily small for sufficiently large $n$ if $\abs{\mathcal L} > \max_{\eta \in \mathcal T} 2^{nS(\tilde{\rho}^\eta)}$. Note that the states $\tilde{\rho}_{\eta}^n, \tilde{\rho}_{\eta,m}^n$ for any $\eta \in \mathcal{S}_E$ can be obtained from $\tilde{\rho}_{\eta_\text{max}}^n, \tilde{\rho}_{\eta_\text{max},m}^n$, where $\eta_\text{max} = \max_{\eta \in \mathcal S_E} \eta$, through a pure loss channel. Therefore the data processing inequality (Thm 9.2 of \cite{Nielsen_Chuang_2010}) implies that any code with the property $\|\tilde{\rho}_{\eta_\text{max}}^n-\tilde{\rho}_{\eta_\text{max},m}^n||_{1} \leq 30\varepsilon^{\frac{1}{4}} + \frac{1}{n}$ has the same property for all other $\eta \in \mathcal S_E$.
The number of actual messages that the sender can privately transmit is therefore upper bounded by
\begin{align}
    \abs{\mathcal M} \leq \min_{\eta \in \mathcal T} \, 2^{n(S(\Bar{\rho})-S(\tilde{\rho}^\eta))}.
\end{align}
One can get arbitrarily close to this bound for sufficiently large $n$ and we get the rate
\begin{align}
    R = S(\Bar{\rho})- \max_{\eta \in \mathcal T} \, S(\tilde{\rho}^\eta)
\end{align}

For the specific case of the \gls{bpsk} ensemble, all we need to do is specify the entropy of the average state, which for this ensemble of pure state is equal to the Holevo quantity, given in lemma \ref{thm:bpsk_chi}.

One aspect that we have not addressed yet is derandomization, or the question of how we go from a random code to a specific code. Since both the data transmission and privacy criteria are fulfilled with high probability for a random code, the existence of a code with both properties is ensured.
\section{Conclusion}
In this work, we derived a capacity formula for the bosonic compound wiretap channel under the assumption of \gls{bpsk} modulation. Our results demonstrate that despite the inherent uncertainty in channel parameters, secure communication can be achieved by leveraging the structural differences between the legitimate and eavesdropper channels. We apply our findings to a satellite communication scenario, illustrating how channel uncertainty affects physical layer security in a real-world setting. Future work could explore more general encoding schemes and relax the assumptions on channel knowledge.

\bibliographystyle{plain}
\bibliography{bib}

\end{document}